\documentclass[review]{elsarticle}
\biboptions{authoryear}
\usepackage{tabularx} 
\usepackage{threeparttable}
\usepackage{microtype}
\usepackage{array}
\usepackage{mathrsfs}
\usepackage{amsmath,bm}
\allowdisplaybreaks[1]
\usepackage{bbm}
\usepackage{amsthm}
\usepackage{appendix}
 \usepackage{amssymb}
\usepackage{float}
\usepackage[font=small,labelfont=bf,tableposition=top]{caption}
\usepackage{booktabs} 
\usepackage{multirow}
\usepackage{amsmath}  
\usepackage[margin=1in,letterpaper]{geometry} 
\usepackage{setspace}
\usepackage[T1]{fontenc}
\usepackage{lmodern}
\usepackage{graphicx}
\usepackage{caption,subcaption}
 \usepackage{threeparttable}
\usepackage{adjustbox}
\usepackage[graphicx]{realboxes}
\usepackage{rotating}
\usepackage{gensymb}
\usepackage{rotating}
\usepackage{color}
\usepackage[final]{hyperref} 
\hypersetup{
	colorlinks=true,       
	linkcolor=blue,        
	citecolor=blue,        
	filecolor=magenta,     
	urlcolor=blue
}

\begin{document}
	
\title{A novel decomposed-ensemble time series forecasting framework: capturing underlying volatility information}

\author {Zhengtao Gui}

\author[] {Haoyuan Li}
\author[] {Sijie Xu}
 \author[]{Yu Chen\corref{cor1}}
\cortext[cor1]{Corresponding author}
\ead{cyu@ustc.edu.cn}	
\address { Department of Statistics and Finance,  School of Management, \\
	University of Science    and Technology of China,  Hefei    230026,   Anhui,
	P. R. China}

\begin{abstract}
Time series forecasting represents a significant and challenging task across various fields. Recently, methods based on mode decomposition have dominated the forecasting of complex time series because of the advantages of capturing local characteristics and extracting intrinsic modes from data. Unfortunately, most models fail to capture the implied volatilities that contain significant information. To enhance the prediction of contemporary diverse and complex time series, we propose a novel time series forecasting paradigm that integrates decomposition with the capability to capture the underlying fluctuation information of the series. In our methodology, we implement the Variational Mode Decomposition algorithm to decompose the time series into $K$ distinct sub-modes. Following this decomposition, we apply the Generalized Autoregressive Conditional Heteroskedasticity (GARCH) model to extract the volatility information in these sub-modes. Subsequently, both the numerical data and the volatility information for each sub-mode are harnessed to train a neural network. This network is adept at predicting the information of the sub-modes, and we aggregate the predictions of all sub-modes to generate the final output. By integrating econometric and artificial intelligence methods, and taking into account both the numerical and volatility information of the time series, our proposed framework demonstrates superior performance in time series forecasting, as evidenced by the significant decrease in MSE, RMSE, and MAPE in our comparative experimental results.

\end{abstract}
\maketitle

 \vskip 0.5cm

\newpage
\section{Introduction}
\label{sec:intro}

Time series forecasting problems have grown increasingly prevalent across diverse fields. The complex attributes inherent in a time series, including seasonality, trend, and level, pose significant challenges in generating accurate forecasts. Concerning seasonality, a time series may display complex behaviors such as multiple seasonal patterns, non-integer seasonality, and calendar effects, among others. Traditional time series models, such as ARIMA\citep{box1970distribution} and GARCH\citep{bollerslev1987conditionally}, rest on a pivotal assumption: the ability to accurately represent data in a linear form. However, this assumption does not hold in practical scenarios. Consequently, researchers have turned to more advanced techniques to handle such complexities in time series data.
 
In recent times, a variety of deep learning approaches based on Artificial Neural Network (ANN) models have been adopted for time series forecasting. Extensive comparative studies have assessed the efficacy of ANNs relative to classical forecasting methods like ARIMA and GARCH. Hybrid models that merge ANNs with GARCH-type models have shown superior performance compared to standalone ANNs or traditional time series models. For example, Bildirici and Ersin \citep{bildirici2013forecasting} combined GARCH with neural networks to forecast oil prices, finding that the hybrid model was promising. Hu et al. \citep{hu2020hybrid} investigated integrating deep learning with the GARCH model to better predict copper price volatility, discovering that GARCH-based forecasts enhanced the volatility predictions when used as features. Furthermore, the incorporation of Recurrent Neural Networks (RNNs), such as LSTM and BLSTM, into the combined GARCH-ANN framework significantly boosted the accuracy of volatility forecasts. These advancements encourage further research and application of deep learning techniques in time series forecasting challenges.

In the realm of neural networks, the recent surge of success is primarily attributed to Recurrent Neural Networks (RNNs) \citep{medsker2001recurrent} and Long Short-Term Memory Networks (LSTMs) \citep{hochreiter1997long}, as they are intrinsically well-suited to modeling sequence data. These models have distinguished themselves due to their remarkable performance in time series prediction. Particularly, LSTM effectively transmits and represents information in long-term sequences, thereby overcoming the problem of neglecting valuable historical information. Moreover, LSTM can efficiently mitigate the vanishing or exploding gradient problem. These neural networks exhibit varying performances in processing distinct datasets and have become prevalent tools utilized by enterprises for forecasting time series.

While the strong performance of neural networks in time series forecasting is well-recognized \citep{zhao2022k,cao2019financial,zhang2023spatial,li2020hierarchical,yadav2023noa}, it does not obviate the relevance of traditional time series models. Volatility, a crucial element in financial time series forecasting, has implications for asset pricing, portfolio selection, and risk management. The GARCH model provides a valuable tool for modeling the fluctuation information inherent in dynamic time series. By introducing such fluctuation information into the neural network as supplementary data, it is possible to enhance forecasting performance \citep{hu2020hybrid}. However, an additional challenge arises: due to their complexity and nonlinearity, raw time series data may be too intricate for neural networks to process effectively, which can hinder neural networks from predicting as accurately as anticipated.

To tackle the aforementioned issue, we propose employing a 'divide and conquer' approach, which involves decomposing the data into sub-modes with distinct characteristics\citep{huang2021hybrid,jianwei2019novel,lin2020crude,fang2023sentiment,zhang2023oil}. These sub-modes can then be individually analyzed using different models. For addressing non-linear and non-stationary problems, a data pre-processing technique called Variational Mode Decomposition (VMD) \citep{dragomiretskiy2013variational} has been demonstrated to be effective in analyzing non-linear and non-smooth datasets. Several studies have applied the VMD method to various domains, including financial stock sequences \citep{guo2022forecasts} and carbon price forecasting \citep{huang2021hybrid}, yielding promising results. These investigations suggest that VMD is highly capable of handling non-linear and non-stationary sequences.

In this research, We employ Variational Mode Decomposition to decompose the data into various components, encompassing both low- and high-frequency parts, and utilize the Generalized Auto-Regressive Conditional Heteroskedasticity model to extract the fluctuation information of each sub-mode. The neural network is subsequently supplied with both the values and variances of the time series as inputs. In addition, a distinct model is trained for each sub-mode to conduct predictions, and the values of each sub-mode are aggregated to generate the final output. This hybrid approach, combining artificial intelligence and econometric methods, has demonstrated greater efficacy than applying either individually. The efficacy of our hybrid model is attributable to the GARCH framework's proficiency in characterizing the stochastic fluctuations and volatility clustering typical of time series data, thereby enriching the analysis with nuances potentially overlooked by neural networks. Concurrently, the neural network demonstrates an exceptional capacity for managing data with long-range dependencies. The decomposition of data amplifies the discernibility of sequence patterns, thus ensuring that each model contributes a distinct and instrumental perspective to the overall predictive accuracy.

The organization of this paper is as follows: Section 2 introduces the models employed in this study and elaborates on the methodology used. Section 3 presents the experiment and corresponding results, and finally, the last section provides a conclusive discussion of our work.

\section{Methodology}

\subsection{Evaluation Metrics}

To assess the predictive accuracy of the models, we utilize three widely recognized metrics for evaluating prediction errors: Root Mean Square Error (RMSE), Mean Absolute Error (MAE), and Mean Absolute Percentage Error (MAPE). The formulas for RMSE, MAE, and MAPE are defined as follows:

\begin{equation}
\operatorname{R M S E}=\sqrt{\frac{\sum_i^n[x(t)-x(\hat{t})]^2}{n}}
\end{equation}
\begin{equation}
\operatorname{M A E}=\frac{\sum_i^n|x(t)-x(\hat{t})|}{n} 
\end{equation}
\begin{equation}
\operatorname{M A P E}= \frac{100\%}{n}\sum_i^n  \Big|\frac{(x(t)-x(\hat{t}))}{x(t)}\Big|
\end{equation} 

Where $x(t)$ and $x(\hat{t})$ represent the actual value and the predicted value. It should be noted that these metrics introduced above, namely MAE, RMSE, and MAPE, are widely referenced in numerous studies. For instance, Hu et al. \cite{hu2020hybrid} utilized MAE, RMSE, and MAPE to assess the forecast results of copper price volatility, while Kristjanpoller and Minutolo \cite{kristjanpoller2014volatility} employed these three measures to examine the volatility prediction results of hybrid neural network models for three Latin-American stock exchange indexes. In line with these studies, we also adopt these three criteria to evaluate the predictive performance of different models.

\subsection{Variational Mode Decomposition}

The Variational Mode Decomposition (VMD) algorithm, an improvement upon the Empirical Mode Decomposition (EMD), was proposed by Dragomiretskiy \cite{dragomiretskiy2013variational}. Conventional decomposition methodologies exhibit stringent constraints regarding their data applicability. Specifically, Fourier transformation is primarily applicable to smooth, periodic signals. On the other hand, wavelet analysis is adept at capturing transient effects within signals, operating under the presumption of non-stationarity and linearity in the data. These traditional techniques highlight the critical need for more flexible and comprehensive methods in analyzing intricate signal structures. By contrast, with good flexibility and adaptability, EMD can capture complex signal features without a prior information. However, EMD has several notable shortcomings, including the issues of pattern mixing, endpoint effects, and the challenge of defining precise stopping conditions. To address these issues, Variational Mode Decomposition (VMD) was introduced. VMD is an adaptive and entirely non-recursive approach to modal decomposition and signal processing. It is typically employed to decompose a one-dimensional input signal into a pre-specified number, denoted as $K$, of Intrinsic Mode Functions (IMFs)\cite{rilling2003empirical}.

Contrasting with the recursive decomposition mode utilized by the Empirical Mode Decomposition (EMD), the Variational Mode Decomposition (VMD) introduces a variational decomposition mode, which fundamentally operates as a Multi-adaptive Wiener Filter Bank. VMD is capable of achieving adaptive segmentation for each component within the frequency domain of a signal. This approach can effectively address the issue of pattern overlap that arises during the EMD decomposition process. Moreover, compared to EMD, VMD demonstrates superior noise robustness and weaker endpoint effects. The process of VMD involves the resolution of variational problems and is composed of two primary stages, which are the formulation and resolution of variational problems. It engages three fundamental concepts: the classical Wiener filter \citep{pratt1972generalized}, the Hilbert transform \citep{johansson1999hilbert}, and frequency mixing. In essence, VMD is formulated as a constrained variational problem:

\begin{equation}{\small 
  \min _{\omega_k, y_k} \sum_{k=1}^K\left\|\partial_t\left[\left(\delta(t)+\frac{i}{t \times \pi}\right) * y_k(t)\right] e^{-i \omega_k t}\right\|_2^2 
\hspace{2em}   \text { subject to } \sum_{k=1}^K y_k=y \label{1}}
\end{equation}
where $y$ is the signal to be decomposed. And $\delta$ is the Dirac distribution, $\mathrm{K}$ is the number of modes, $*$ is convolution and ${i}$ is $\sqrt{-1}$.

By utilizing the Augmented Lagrangian Method \citep{fortin2000augmented} (where $\lambda(t)$ represents the Lagrangian multiplier), the initial problem is transformed into an unconstrained variational problem:
\begin{equation}\small
L\left(y_k, \omega_k, \lambda\right)=\alpha \sum_{k=1}^K\Big\|\partial_t\Big[\Big(\delta(t)+\frac{i}{t \times \pi}\Big) * y_k(t)\Big] e^{-i \omega_k t}\Big\|_2^2+\Big\|y(t)-\sum_{k=1}^K y_k(t)\Big\|_2^2+\Big\langle\lambda(t), y(t)-\sum_{k=1}^K y_k\Big\rangle\label{2}
\end{equation}

Where \( \alpha \) is the balance parameter that determines the completeness of the decomposition, the VMD method's decomposition completeness can be fine-tuned by selecting an appropriate \( \alpha \). Subsequently, we apply the Alternating Direction Method of Multipliers (ADMM) \citep{ghadimi2014optimal} to solve the variational problem described. Through iterative alternate update calculations, values for $y_k$ and $\omega_k$ can be obtained.

\subsection{GARCH Model}

Bollerslev's Generalized Autoregressive Conditional Heteroscedasticity model (GARCH) is a classic model to describe cluster volatility. GARCH models are widely utilized in the field of econometrics to predict future levels of volatility based on past observations of the variable itself and past volatilities. For a sequence $\left\{a_t\right\}$ that conforms to GARCH $(k, l)$, it conforms to the following characteristics:

\begin{equation}
a_t=\sigma_t \xi_t
\end{equation}
\begin{equation}
\quad\sigma_t^2=\alpha_0+\sum_{i=1}^k \alpha_i a_{t-i}^2+\sum_{j=1}^l \beta_j \sigma_{t-j}^2
\end{equation}
\begin{equation}
\text{subject to } \alpha_0 > 0, \alpha_i \geq 0, \beta_j \geq 0, 0 < \sum_{i=1}^{k} \alpha_i + \sum_{j=1}^{l} \beta_j \leq 1
\end{equation}

where $\alpha_0$ is a constant coefficient, and $\sigma_t$ is the volatility, which is the conditional standard deviation of returns. $\{\xi_t\}$ denotes an independent identically distributed white noise column with zero mean and unit variance. It is important to note that $\{a_t\}$ represents a stationary sequence and exhibits significant volatility around a mean value. The high-frequency sequence produced by the VMD model manifests robust volatility and, due to its brief period, lacks a conspicuous trend. As such, its sequence characteristics align closely with the GARCH model. Given the stability of the high-frequency series' volatility model, its variation characteristics can be adequately described using the GARCH model.

\subsection{Neural Network}

\subsubsection{Classical RNN Model}

Recurrent Neural Networks \citep{zaremba2014recurrent} are a class of neural networks that excel in processing sequential data. They are particularly well-suited for tasks where the context from earlier in the sequence is informative for understanding the current element, such as in language modeling or time series prediction. 

The fundamental feature of RNNs is their internal state, which captures information about previous elements in the sequence. At each step in the sequence, the RNN updates its state using both the new input and the previous state. This process can be described by the following equations:

\begin{equation}
    h_t = \phi(W_{hh} h_{t-1} + W_{xh} x_t + b_h)
\end{equation}
\begin{equation}
    y_t = W_{hy} h_t + b_y
\end{equation}

Here, $x_t$ is the input vector at time step $t$, $h_t$ is the hidden state vector at time step $t$, $y_t$ is the output vector at time step $t$, and $W_{hh}$, $W_{xh}$, and $W_{hy}$ are weight matrices for the connections between the hidden state-to-hidden state, input-to-hidden state, and hidden state-to-output, respectively. The biases are denoted by $b_h$ and $b_y$, and $\phi$ represents the activation function, typically a non-linear function like tanh or ReLU.

The RNN's ability to maintain a 'memory' through its hidden states allows it to make use of past context. However, standard RNNs can struggle with long-term dependencies due to the vanishing gradient problem, where the contribution of information decays geometrically over time. This limitation has led to the development of more sophisticated RNN architectures such as Long Short-Term Memory (LSTM) networks and Gated Recurrent Unit (GRU) networks, which are designed to better capture long-range dependencies in data.

\subsubsection{GRU Model}

Gated Recurrent Units \citep{cho2014properties} are a type of recurrent neural network architecture that aims to solve the vanishing gradient problem encountered by traditional RNNs during the training process. GRUs achieve this by introducing gating mechanisms that regulate the flow of information. These gates determine how much of the past information needs to be passed along to the future, thus allowing the network to retain long-term dependencies.

The GRU modifies the standard recurrent unit by incorporating two gates: the update gate ($z_t$) and the reset gate ($r_t$). These gates are vectors that decide which information will be passed to the output. The GRU’s update equations are as follows:

\begin{equation}
    z_t = \sigma(W_z \cdot [h_{t-1}, x_t])
\end{equation}
\begin{equation}
    r_t = \sigma(W_r \cdot [h_{t-1}, x_t])
\end{equation}
\begin{equation}
    \tilde{h}_t = \tanh(W \cdot [r_t \ast h_{t-1}, x_t])
\end{equation}
\begin{equation}
    h_t = (1 - z_t) \ast h_{t-1} + z_t \ast \tilde{h}_t
\end{equation}

Here, $\sigma$ represents the sigmoid function, and $\ast$ denotes element-wise multiplication. $W_z$, $W_r$, and $W$ are parameter matrices that need to be learned during training. The vectors $z_t$ and $r_t$ are the update and reset gates, respectively, $h_{t-1}$ is the previous hidden state, $x_t$ is the input vector at time step $t$, $\tilde{h}_t$ is the candidate hidden state, and $h_t$ is the current hidden state of the network.

By carefully balancing new information with past memory, the GRU is able to overcome issues with training over long sequences, making it an effective architecture for tasks requiring the modeling of temporal dynamics.

\subsubsection{LSTM Model}

Long Short Term Memory \citep{yu2019review}, introduced by Hochreiter and Schmidhuber, is a distinctive variant of Recurrent Neural Networks. It was specifically designed to counter the challenge of long-term information forgetting during RNN training, thereby making it apt for describing the long-term influence present in low-frequency sub-modes.
\begin{figure}[H]
    \centering
    \includegraphics[width=0.8\textwidth,height=0.4\textwidth]{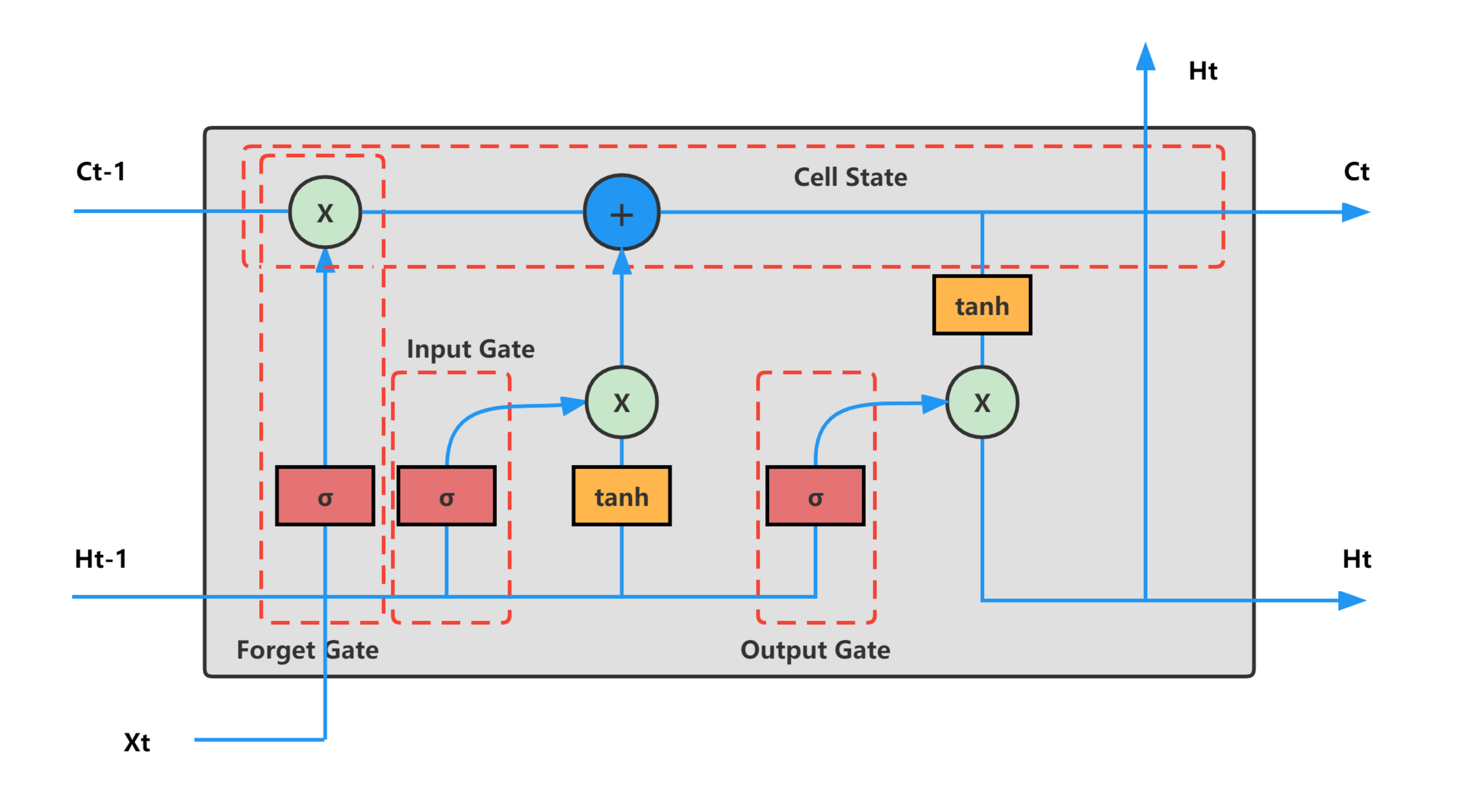}
    \caption{Illustrative diagram of LSTM}
    \end{figure} 
As shown in Figure 1, there are two hidden states $\left(h_t\right.$ and $c_t$) and three gates in LSTM. The first gate is forget gate $f_t$, determining how much information in last hidden state $c_{t-1}$ is to be forgotten, while $f_t$ is decided by an input $x_t$, last hidden state $h_{t-1}$, and a parameter $b_f$ :
\begin{equation}
f_t=\sigma\left(W_f\left[h_{t-1}, x_t\right]+b_f\right) \label{4}
\end{equation}
where $\sigma$ is a sigmoid function to make sure $f_t$ (and the other two gates) is a coefficient between 0 and  1: ~$0$ means complete oblivion and 1 means no information is discarded.
The second gate is input gate $i_t$, and a candidate vector for hidden state $\tilde{C}_t$ :
\begin{equation}
i_t =\sigma\left(W_i\left[h_{t-1}, x_t\right]+b_i\right) \hspace{3em} 
\tilde{C}_t =\tanh \left(W_c\left[h_{t-1}, x_t\right]+b_c\right) \label{5}
\end{equation}
The value domain of $tanh$ is $[-1,1]$, and the new state $C_t$ is a linear combination of the state of the present unit and the previous unit:
\begin{equation}
C_t=f_t * C_{t-1}+i_t * \tilde{C}_t \label{6}
\end{equation}
The last gate is output gate $o_t\left(b_i, b_c\right.$ and $b_o$ are parameters):
\begin{equation}
o_t=\sigma\left(W_o\left[h_{t-1}, x_t\right]+b_o\right) \label{7}
\end{equation}
Output $y_t$ and next hidden state $h_t$ are determined by:
\begin{equation}
y_t=h_t=o_t * \tanh \left(C_t\right)  \label{8}
\end{equation}

\subsection{The Decomposed-ensemble Time Series Forecasting Framework}

\subsubsection{Time Series Decomposition}

Initially, we apply the VMD to decompose the time series into $K$ sub-modes, arranged from high frequency to low frequency. The selection of the number of modes subject to decomposition is performed through judgment rather than algorithmic derivation. Consequently, we utilize the modal quantities ascertained from the EMD process. As illustrated in Figure 2, the VMD yields ten distinct sub-modes from the decomposed time series. These sub-modes are characteristically independent, displaying a clear and orderly structure with stable and consistent fluctuations, thereby streamlining our modeling process. For the purpose of analytical clarity, these modes are bifurcated into two classifications predicated on their frequency attributes:

1. The high-frequency component, characterized by high frequency and low amplitude, shows frequent short-term fluctuations but also holds long-term significance. A tendency for major fluctuations to precede and succeed each other, as well as for minor fluctuations, suggests the presence of volatility clustering.

2. The low-frequency component, with its high amplitude, reflects changes influenced by external factors. Significant events may cause this component to quickly rise or fall. While its fluctuation frequency is low, it can bring about substantial changes in the time series or even shift its underlying mechanism.

From the analysis above, different frequency components exhibit unique data characteristics. By adopting forecasting models tailored to each specific component, we can enhance the accuracy of our predictions.

\begin{figure}[H]
    \centering
    \includegraphics[width=0.45\textwidth,height=0.45\textwidth]{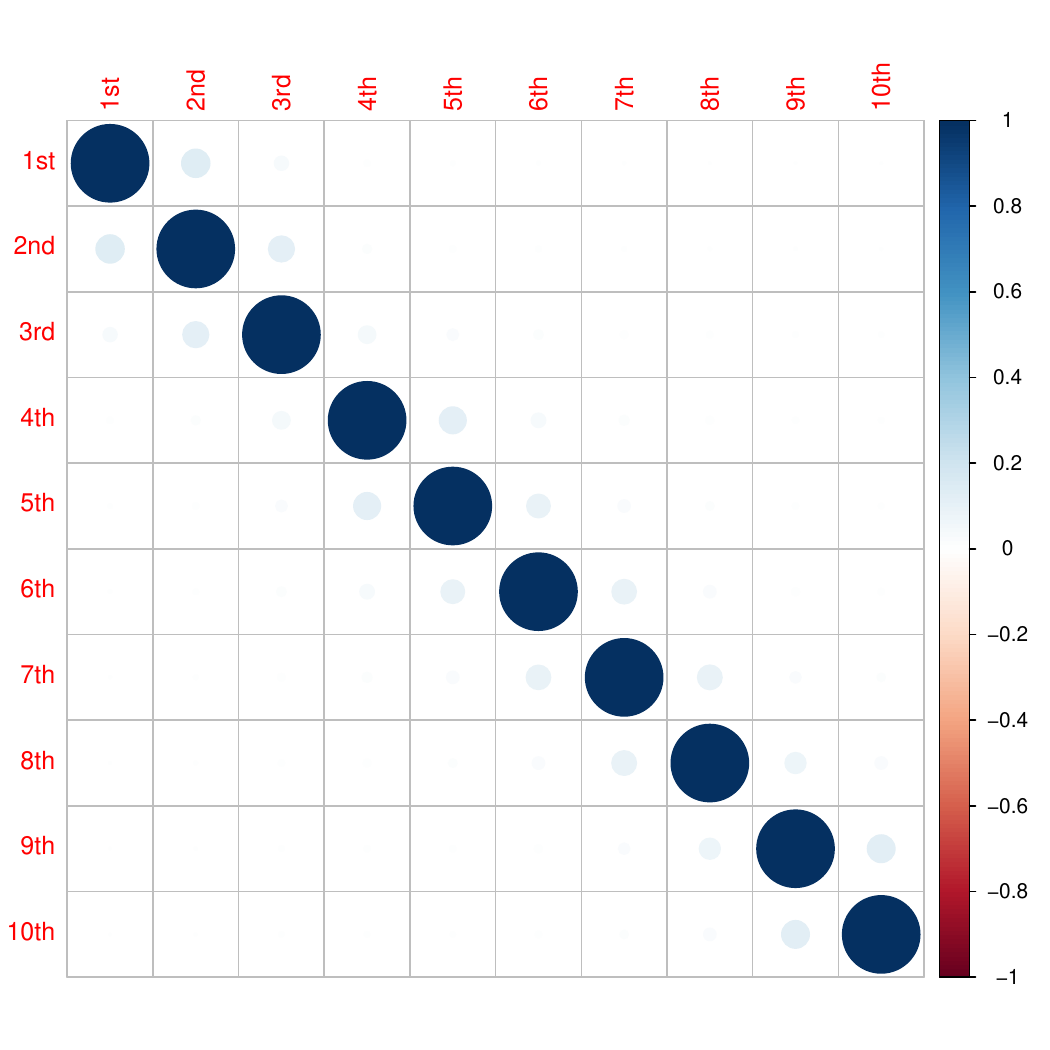}
    \caption{The correlation plot of the 10 decomposed series}
    \end{figure} 

\subsubsection{Using GARCH Model to Extract Volatility Information of Sub-modes}

Secondarily, we conduct the ADF test on the GARCH models associated with each sub-modes to evaluate their model validity, if the residuals of the model fit are stationary, the GARCH model will be constructed. The fluctuation information extracted by the GARCH model will be used as the input of the neural network. Time series data typically exhibit a mix of short-term and long-term fluctuations. Figure 3 illustrates the variance sequence of three high-frequency sub-modes derived from a VMD-decomposed time series. The GARCH model excels at extracting such fluctuation data, which is likely to enhance the forecasting capability of neural network. Moreover, it is significantly easier to extract information from these decomposed sub-modes than from the raw time series data.

\begin{figure}[H]
    \centering
    \includegraphics[width=0.8\textwidth,height=0.25\textwidth]{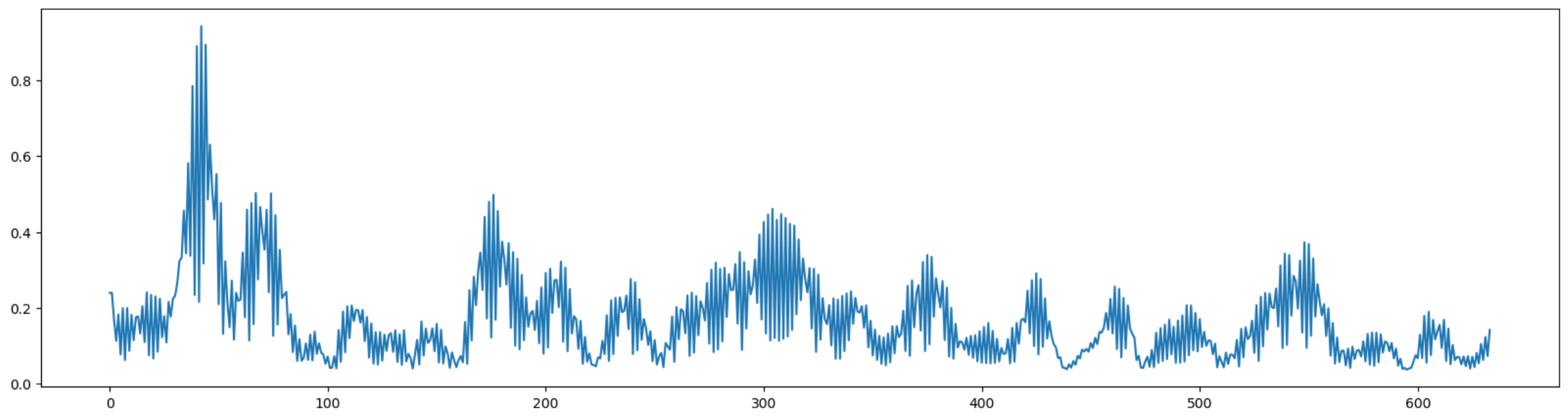}
    \end{figure} 
\begin{figure}[H]
    \centering
    \includegraphics[width=0.8\textwidth,height=0.25\textwidth]{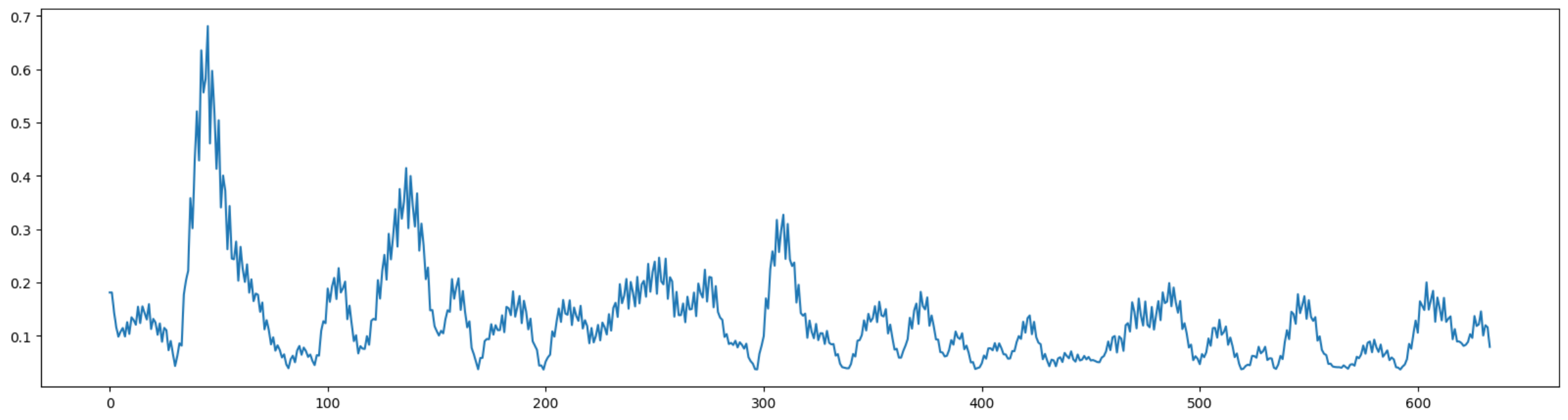}
    \end{figure} 
\begin{figure}[H]
    \centering
    \includegraphics[width=0.8\textwidth,height=0.25\textwidth]{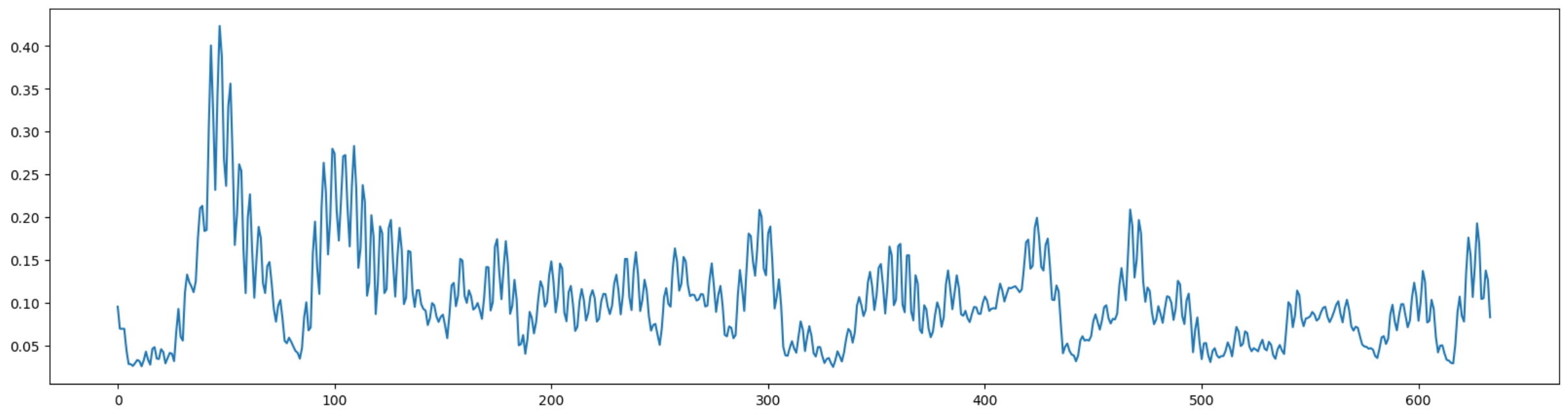}
    \caption{The volatilities of the three high-frequency sequence sub-modes}
    \end{figure} 

\subsubsection{Input the Numerical and Volatility Information of Each Sub-mode to Train the Neural Network} 

This procedure entails inputting the sequence value of each sub-mode and the variance information derived from GARCH model into the neural networks for training. Each sub-mode is subsequently predicted using these neural networks, and the predicted values of all sub-modes are added together to obtain the final output. Training the neural network on each sub-mode, rather than directly on the time series, improves the model's overall performance since the sub-modes offer more information for the neural network to learn from. To ensure the comparability of the models, all neural network models are configured with identical parameters, which will be further detailed in the experimental section. The proposed multi-scale, nonlinear ensemble learning approach for forecasting time series, termed VMD-GARCH-ANN, adeptly integrates VMD, GARCH, and neural networks. Figure 4 depicts the process of the framework.

\begin{figure}[htbp]
    \centering
    \includegraphics[width=0.8\textwidth, height=0.9\textwidth]{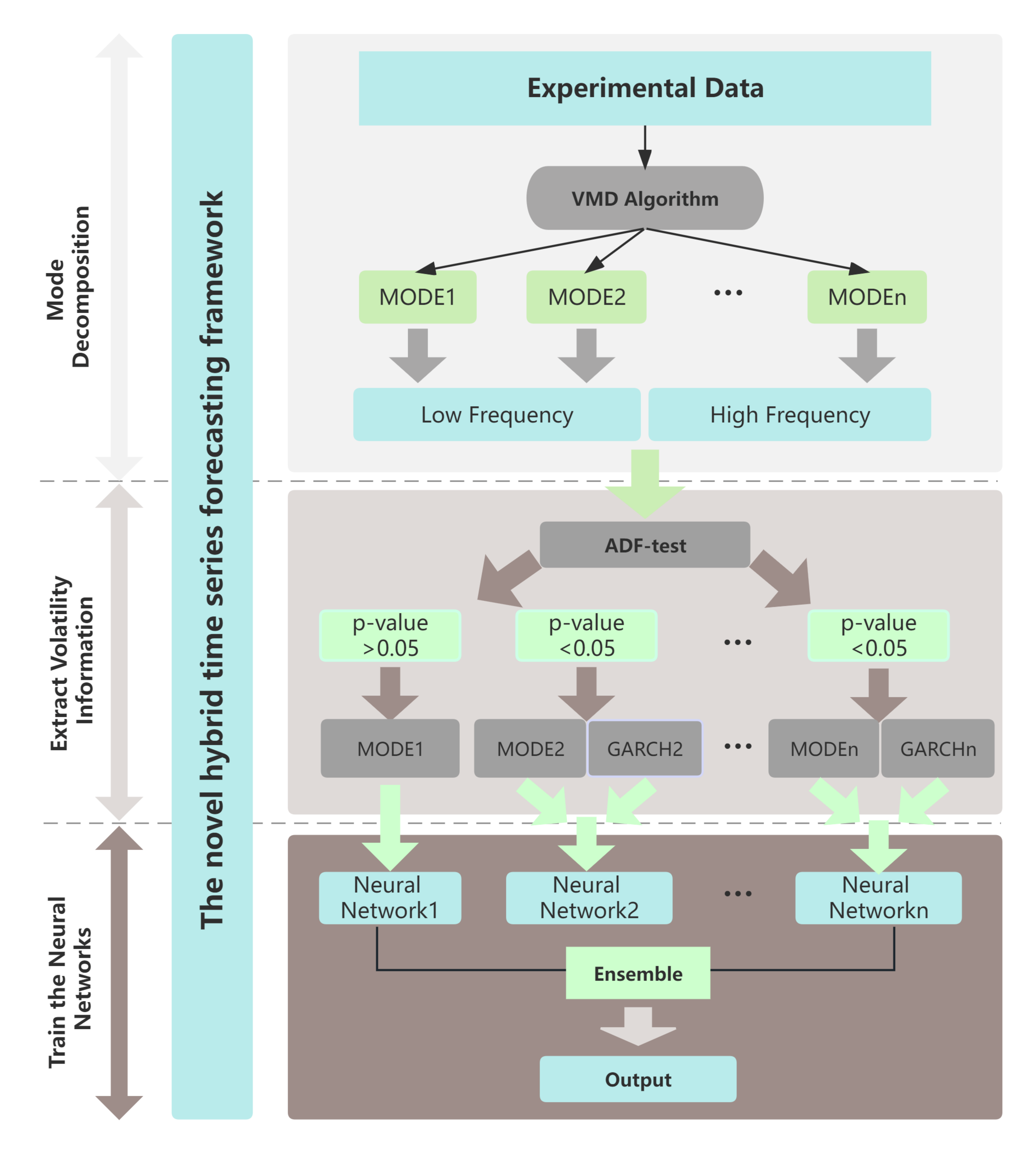}
    \caption{Time-series forecasting process of our framework.}
    \end{figure} 

\section{Experimental Results}

\subsection{Data Introduction}

We implement our method on the Consumer Price Index (CPI) data for Germany, sourced from FRED, which spans from 1970 to 2022 and is recorded monthly. An analysis of the data reveals that apart from minor fluctuations in the CPI under standard economic operations, significant impacts arise from external events and the macroeconomic environment. Notably, in 1973, Germany grappled with the oil crisis, triggering a peak in the CPI. Following the signing of the Plaza Agreement in 1985, the CPI sustained an upward trend until 1995. The data is visually represented in Figure 6.

\begin{figure}[htbp]
    \centering
    \includegraphics[width=0.8\textwidth,height=0.50\textwidth]{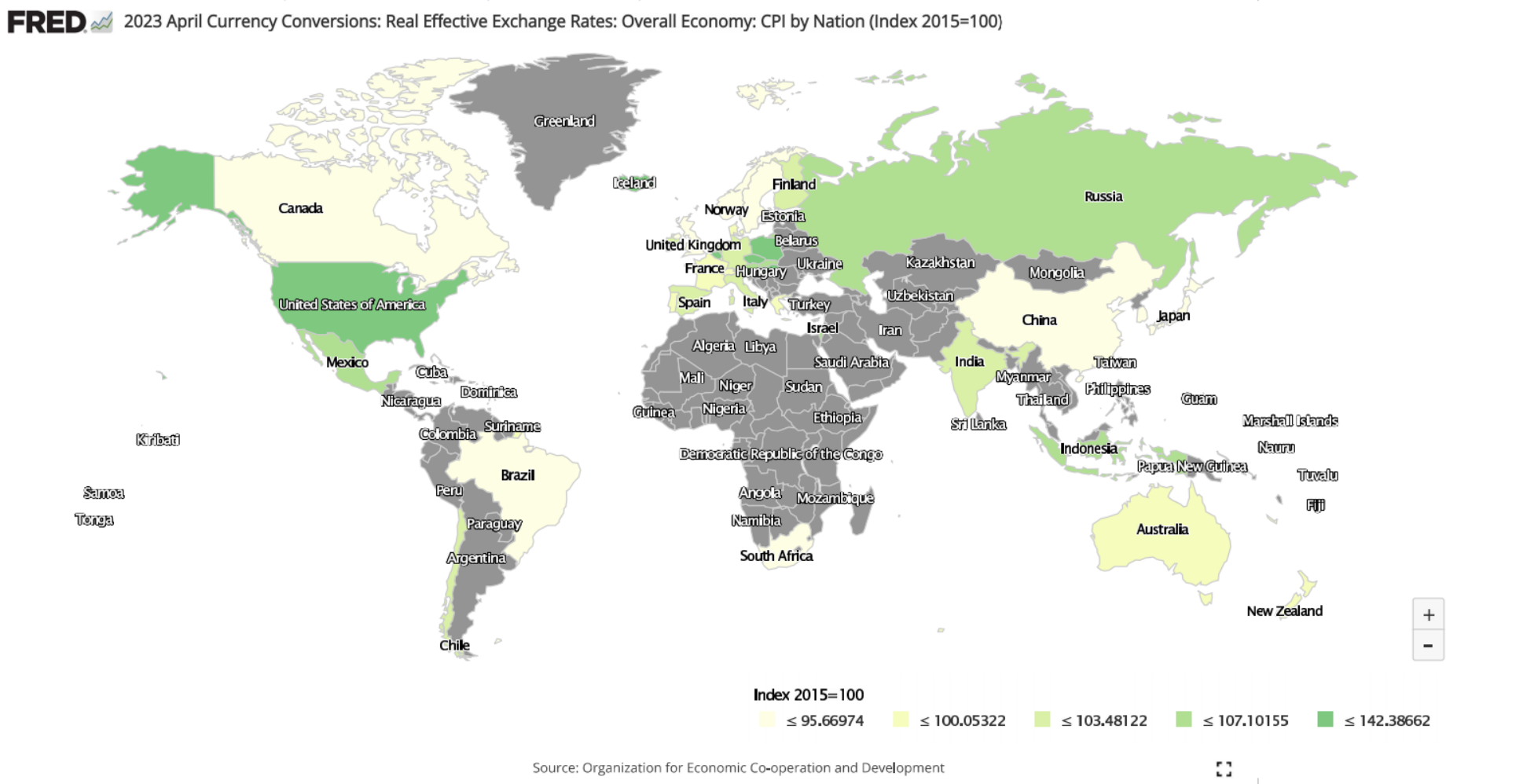}
    \caption{2023 April Currency Conversions: Overall Economy: CPI by Nation(Index 2015=100)}
    \end{figure} 

\begin{figure}[htbp]
    \centering
    \includegraphics[width=0.8\textwidth,height=0.48\textwidth]{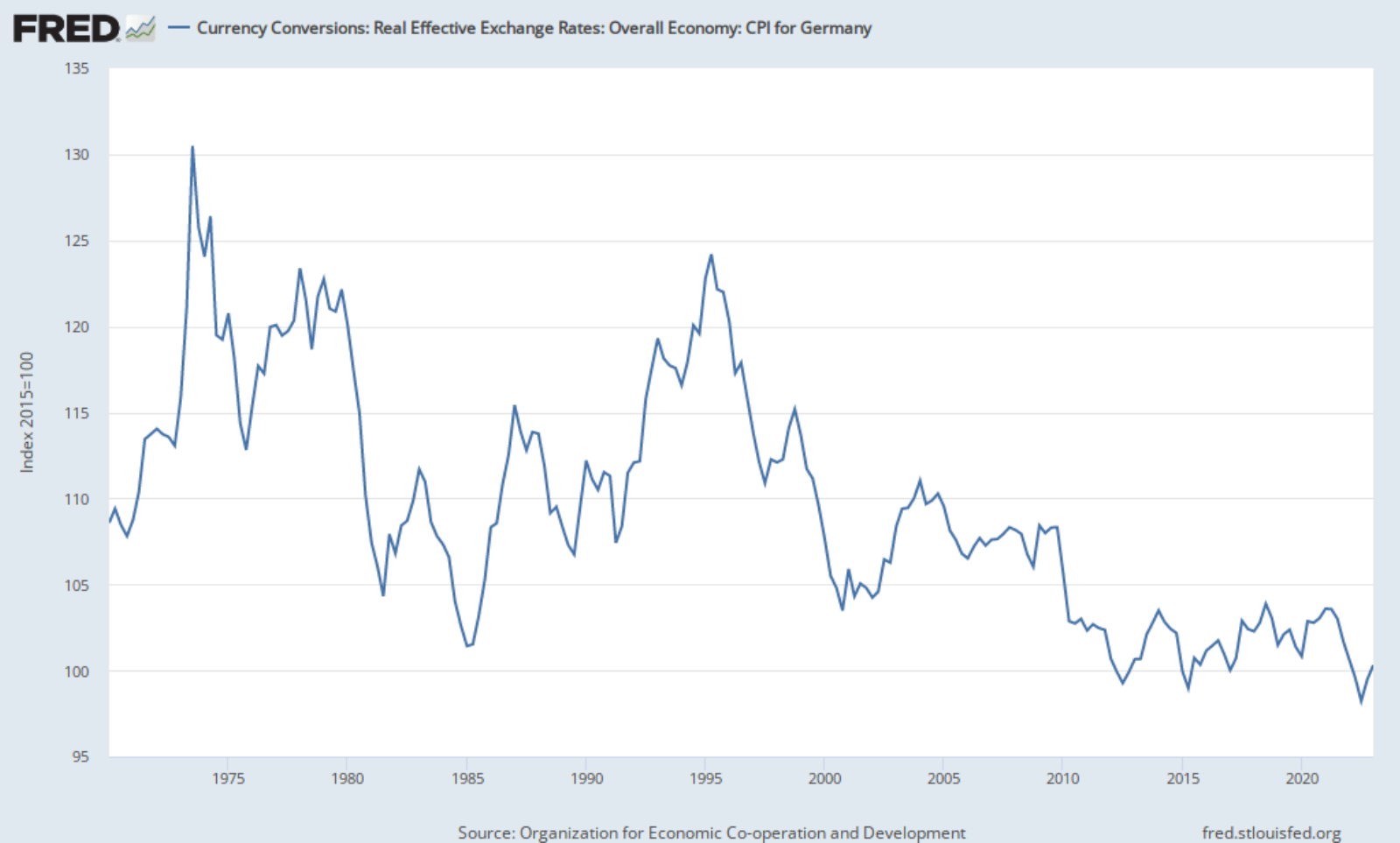}
    \caption{CPI of Germany from 1970 to 2022}
    \end{figure} 

\begin{table}[!h]
    \caption{the attributes of the original CPI data and decomposed series(listed by frequency from low to high)}
    \centering
    \small
    \begin{tabular}{c c c c c c c c}
        \hline
        Series & Mean & Variance & Min & Max & Variance & ADF test & ARCH(12)\\
        \hline
        
        \vspace{1em}
        Original data &109.783	&46.651	&98.296	&132.981	&54.366&	-0.349	&1.60E-122 \\
        \vspace{1em}

        Series1&	109.782	&23.550	&101.621&	117.225	&27.445	&-1.294	&2.12E-125
\\
        \vspace{1em}
        
        Series2&	2.60E-06&	11.460&	-8.095&	7.234	&13.355&	-14.533	&2.12E-125
  \\
        \vspace{1em}
        
        Series3&	1.96E-07&	3.385&	-4.914&	5.446	&3.944&	-41.215&	2.15E-125
\\
        \vspace{1em}
        
Series4&	1.43E-08	&0.399	&-2.248&2.418&	0.465	&-114.631&	7.48E-125

\\
        \vspace{1em}
        
Series5&6.92E-09&	0.196&	-1.718&	1.905&	0.229&	-155.694&	1.83E-123
\\
        \vspace{1em}

Series6&	3.57E-09&	0.071&	-0.708&	0.771&	0.082&	-167.432&	1.03E-114
\\
        \vspace{1em}
        
Series7&	1.97E-09&	0.042&	-0.998&	0.944&	0.049&	-183.151&	1.99E-119
 \\
        \vspace{1em}
        
 Series8	&1.14E-09	&0.026	&-0.845	&0.879	&0.0313&-191.597&	6.08E-118
\\
\vspace{1em}
Series9&	6.59E-10&	0.014&	-0.462&	0.539&	0.017&	-196.237&	1.44E-114

\\

\vspace{1em}
Series10&	4.83E-10&	0.010&	-0.319&	0.315&	0.012	&-155.746	&4.34E-123

\\

\hline
    \end{tabular}
\end{table}

\subsection{Numerical Experiment}

Given the non-stationary and non-linear characteristics of the data, we utilize Variational Mode Decomposition (VMD), an adaptive method that avoids recursion, for decompose the raw data. The high-frequency mode presents pronounced randomness, while the low-frequency mode is distinguished by notable volatility. Additionally, the first mode captures the overarching trend of the Consumer Price Index (CPI). These decomposed modes present a more straightforward structure with more consistent and predictable fluctuations compared to the original CPI series, potentially improving the precision of model fitting and forecasts. The modes are sorted by increasing frequency, with the total number of sub-modes fixed at ten. Table 1 displays the features of the original CPI data alongside the attributes of the decomposed series.

\begin{figure}[htbp]
    \centering
    \includegraphics[width=0.9\textwidth,height=1.2\textwidth]{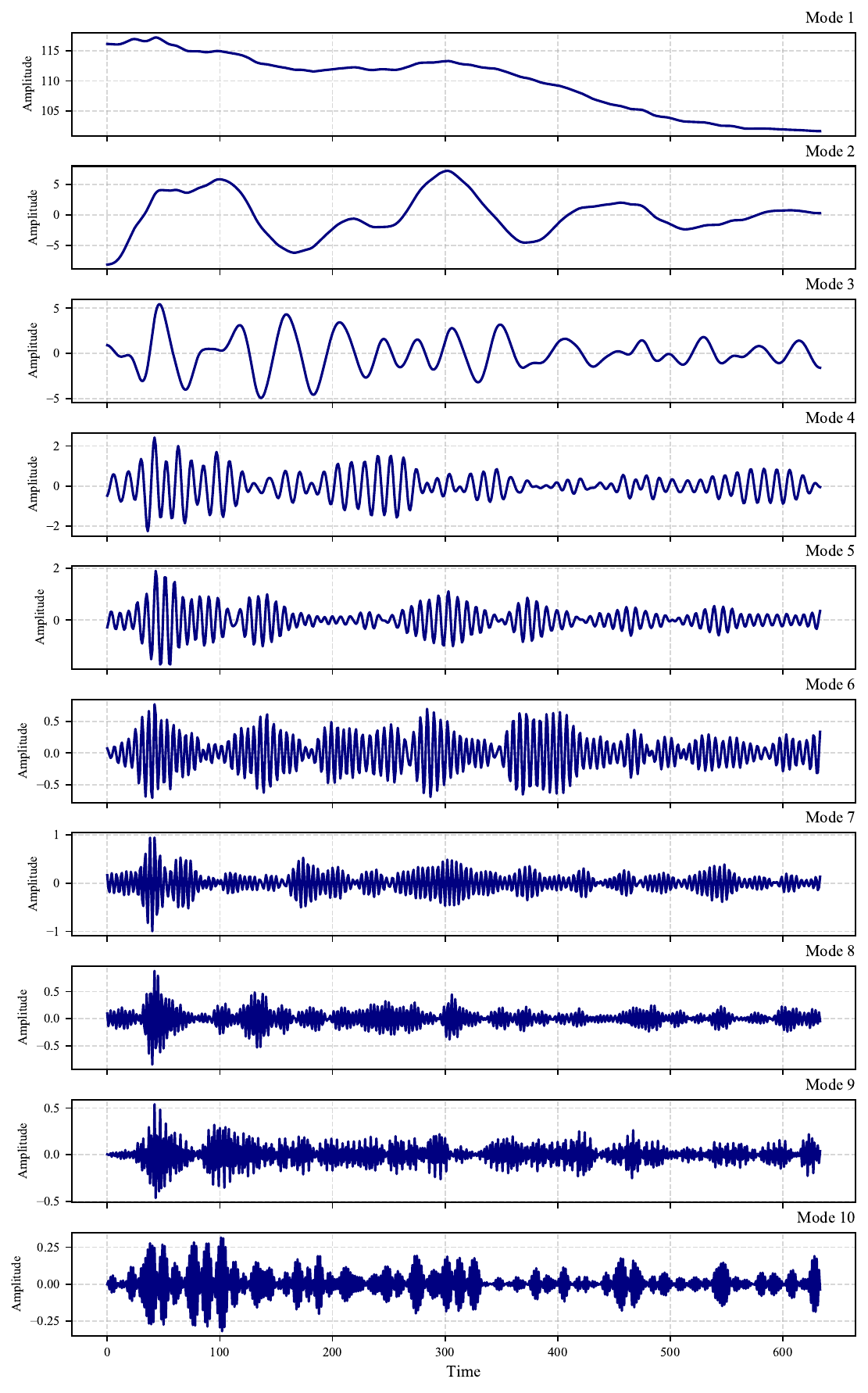}
    \caption{Decomposition result of CPI data}
    \end{figure} 

Figure 7 demonstrates that after filtering out random noise, the low-frequency component of the time series has reduced complexity and reflects the overarching trend. In contrast, the high-frequency component, characterized by lower volatility, does not display a clear long-term trend, but rather transient fluctuations which represent short-term disturbances in the CPI data. This distinction suggests that the CPI possesses unique properties in its high-frequency and low-frequency elements. Short-term CPI fluctuations are primarily driven by immediate economic conditions and exert a minimal effect on the long-term trend. The long-term CPI fluctuations are influenced by broader external factors such as policy changes and global economic shifts.

In this study, we employ GARCH (10, 10) model for each segment, extracting their respective volatility data. To highlight the superior performance of the integrated model, we leverage three types of neural networks for modeling: Classical Recurrent Neural Network (RNN), Gated Recurrent Unit (GRU), and Long Short-Term Memory (LSTM) network. Each neural network variant gives rise to three distinct prediction models: one directly modeled using the neural network, another combined with the VMD decomposition approach, and a third combined with both VMD and GARCH. Consequently, we create a suite of nine different models. To ensure a fair comparison, we unify the neural network model parameters across all training. Each neural network model consists of two layers, followed by a dropout layer \citep{baldi2013understanding} to avert overfitting. Among adaptive techniques, the Adam optimizer \citep{kingma2014adam} is particularly effective and superior to other methods in practical applications. As such, we utilized it to iteratively update the weights of our neural network model based on the training dataset. In the experiment, we employed Mean Squared Error (MSE) as the loss function. We allocated the first 85 percent of the sub-modal sequential values from the dataset to the training dataset, with the remaining 15 percent serving as the testing dataset. By combining 50 steps sequences and their corresponding values from the modal volatility sequence, we crafted a vector with a length of 100 to serve as the input of neural network.

Our experimental design employs a one-step-ahead real-time prediction strategy, which involves appending the actual price of a given day to the training set after making a prediction, thus facilitating subsequent forecasts. To benchmark the performance of various prediction models, we utilize three widely recognized loss functions: root mean square error (RMSE), mean absolute error (MAE), and mean absolute percentage error (MAPE). These metrics are frequently employed in price forecasting research. The comparative results across different models are depicted in Figure 8 and summarized in Table 2. The term 'LSTM' refers to predictions made directly using the LSTM network on the time series, while 'VMD-LSTM' denotes the approach where the time series is first decomposed using VMD, followed by employing LSTM to predict each decomposed sub-mode, then aggregate the prediction results. For our analysis, we selected a sequence of ten consecutive values from the test set for predictions by each model.

Figure 8 presents the out-of-sample prediction results for the aforementioned models. Upon comparison, it becomes evident that irrespective of the neural network utilized, the performance of our integrated model framework surpasses that of both the basic model and the model employing VMD decomposition integration. The predictions made by the VMD-GARCH-LSTM, VMD-GARCH-GRU, and VMD-GARCH-RNN models align more accurately with the actual value. This accuracy is reflected in both magnitude and direction. In other words, there is not only a smaller quantitative discrepancy between the predicted and actual values but also a more consistent upward or downward trend at each time point.

\begin{figure}[]
 \centering
 {
  \begin{minipage}{7cm}
   \centering
   \includegraphics[scale=0.5]{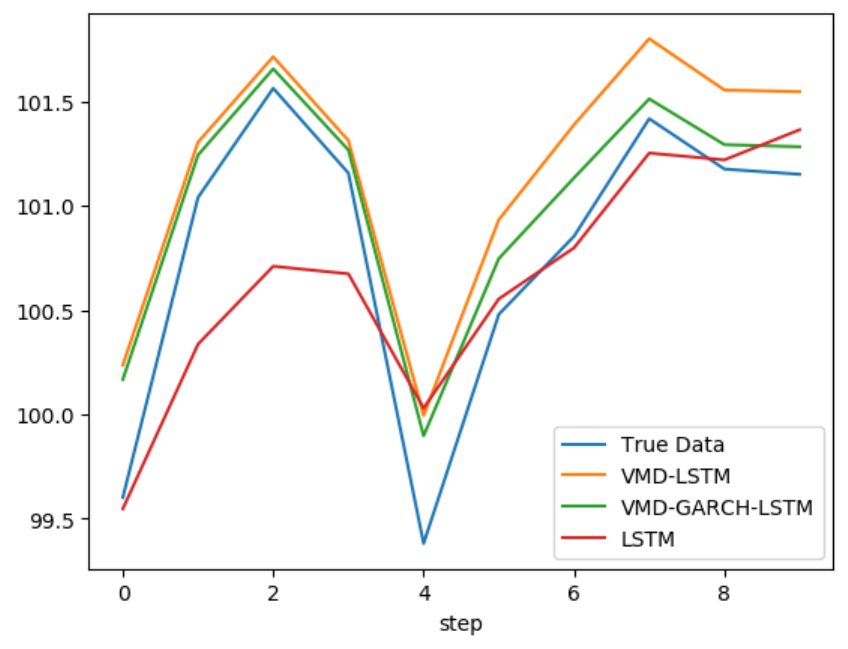}
  \end{minipage}
 }
    {
     \begin{minipage}{7cm}
      \centering
      \includegraphics[scale=0.5]{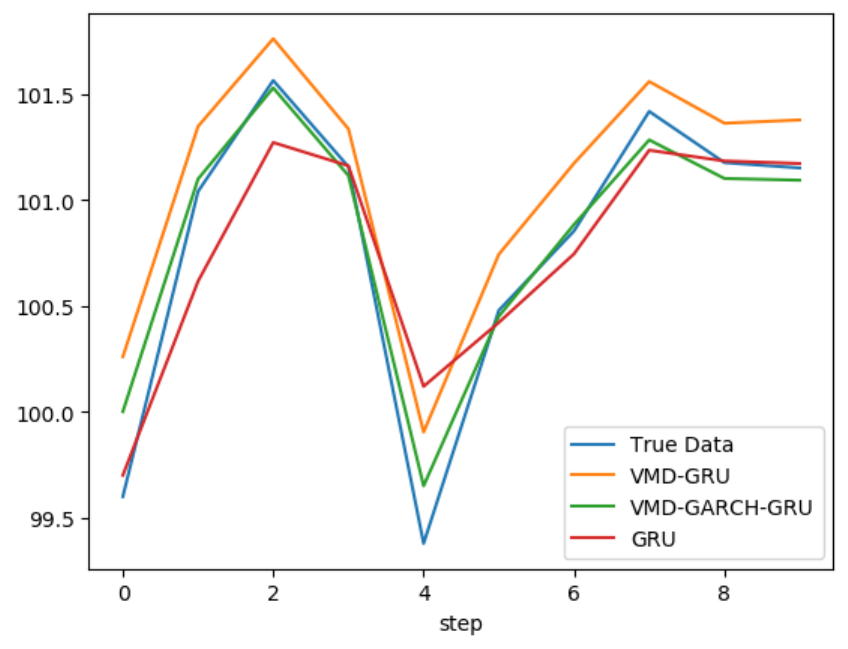}
     \end{minipage}
    }
     {
  \begin{minipage}{7cm}
   \centering
   \includegraphics[scale=0.5]{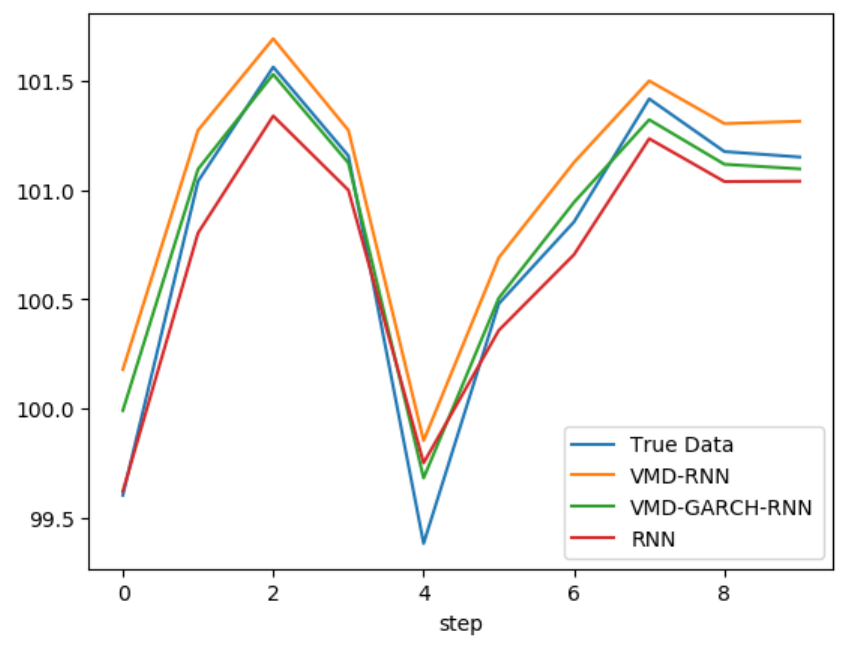}
  \end{minipage}
 }
\caption{\fontsize{33.5bp}{10bp}10-step forecast performance of different models}
\label{fig:14}
\end{figure}

\begin{figure}[htbp]
    \centering
    \includegraphics[width=1\textwidth,height=0.35\textwidth]{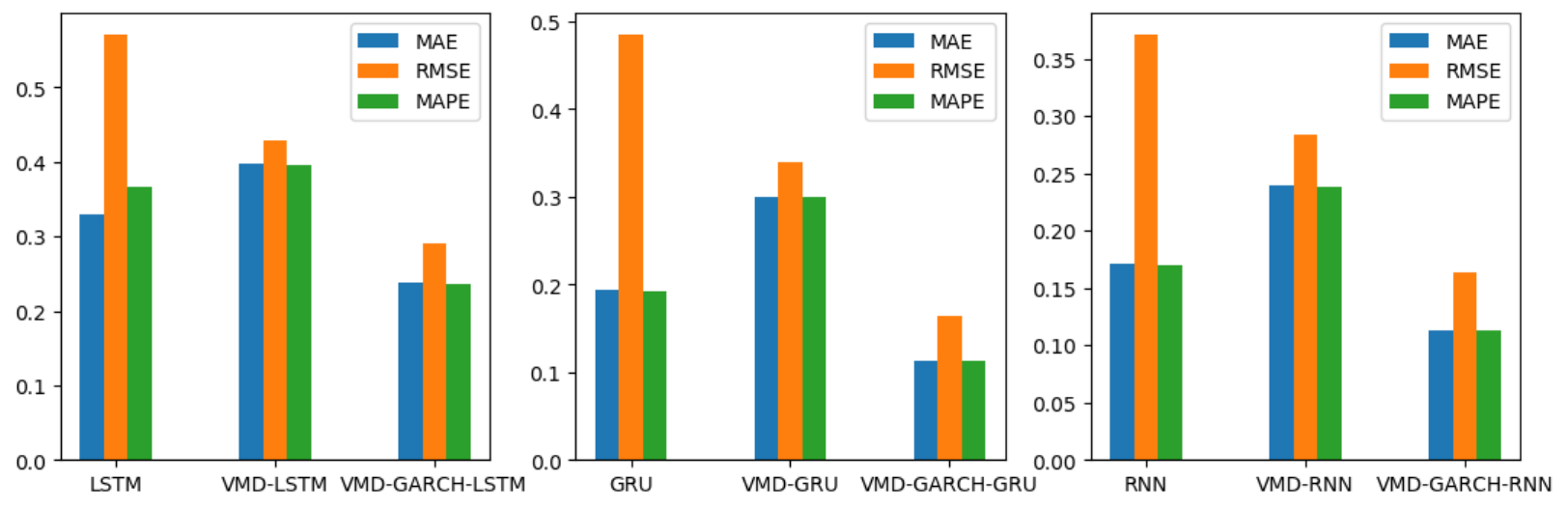}
    \caption{Evaluation metrics for different models}
    \end{figure}

\begin{table}[!h]
    \caption{Performance of different methods under RMSE MAE and MAPE}
    \centering
    \small
    \begin{tabular}{c c c c}
        \hline
        Model/Criteria \hspace{5em} & MAE \hspace{5em} & RMSE &  \hspace{5em} MAPE\\
        
        \hline
        
        \vspace{1em}
        LSTM \hspace{5em} & 0.330\hspace{5em} & 0.571 \hspace{5em}& \hspace{5em} 0.367\% \\
        \vspace{1em}
        GRU \hspace{5em} & 0.194\hspace{5em} & 0.485\hspace{5em}& \hspace{5em} 0.193\%  \\
        \vspace{1em}
        RNN \hspace{5em} & 0.171\hspace{5em} &
        0.371\hspace{5em}& \hspace{5em} 0.170\%  \\
        \vspace{1em}
        VMD-LSTM \hspace{5em} & 0.397\hspace{5em} & 0.429\hspace{5em}& \hspace{5em} 0.395\%  \\
        \vspace{1em}
        VMD-GRU \hspace{5em} & 0.3\hspace{5em} & 0.339\hspace{5em}& \hspace{5em} 0.299\%  \\
        \vspace{1em}
        VMD-RNN \hspace{5em} & 0.239\hspace{5em} & 0.284\hspace{5em}& \hspace{5em} 0.238\%  \\
        \vspace{1em}
        VMD-GARCH-LSTM \hspace{5em} & 0.238\hspace{5em} & 0.290\hspace{5em}& \hspace{5em} 0.237\%  \\
        \vspace{1em}
        VMD-GARCH-GRU \hspace{5em} & 0.113\hspace{5em} & 0.164\hspace{5em}& \hspace{5em} 0.113\%  \\
        VMD-GARCH-RNN \hspace{5em} & 0.113\hspace{5em} & 0.164\hspace{5em}& \hspace{5em} 0.113\%  \\
        \hline
    \end{tabular}
\end{table}

In Figure 9 and Table 2, we apply the three aforementioned evaluation metrics-root mean square error (RMSE), mean absolute error (MAE), and mean absolute percentage error (MAPE)-to assess and compare the predictive performance of the various models. The results from these evaluation indicators further underscore the exceptional predictive performance of the combined model. Modal decomposition transforms a complex consumer price index series into a stationary and regular structure, which significantly enhances the prediction accuracy rate.

In the following experiments, we compared the predictive performance of three models—LSTM, VMD-LSTM, and VMD-GARCH-LSTM—across multiple forecasting intervals. The neural network's model parameter settings are consistent with those employed in previous experiments. To examine the robustness of different models under various circumstances and discern differences in their evaluation metrics, we used six distinct sets of prediction steps.

\begin{figure}[htbp]
    \centering
    \includegraphics[width=0.9\textwidth,height=0.66\textwidth]{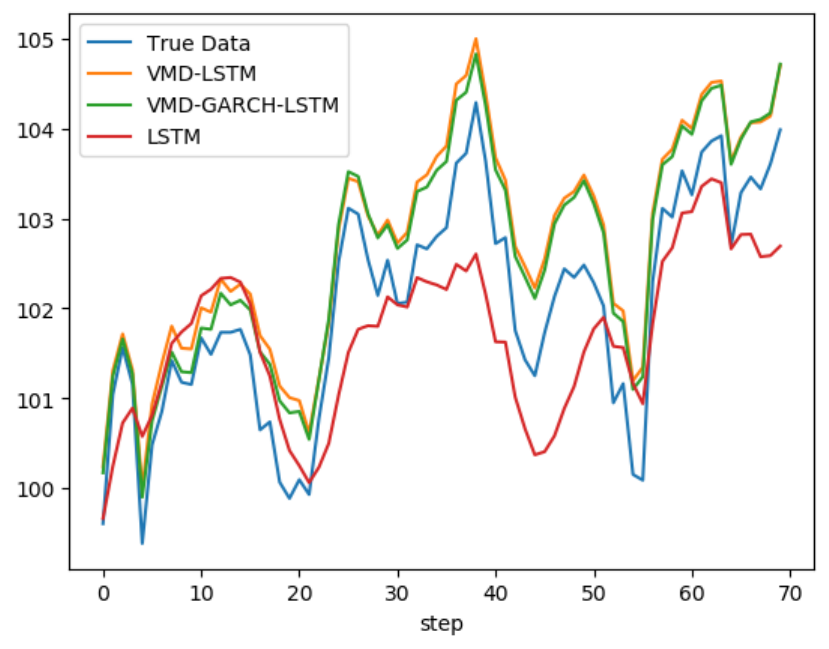}
    \caption{The performance of each model for 70-step prediction}
    \end{figure} 

\begin{figure}[htbp]
    \centering
    \includegraphics[width=1\textwidth,height=0.55\textwidth]{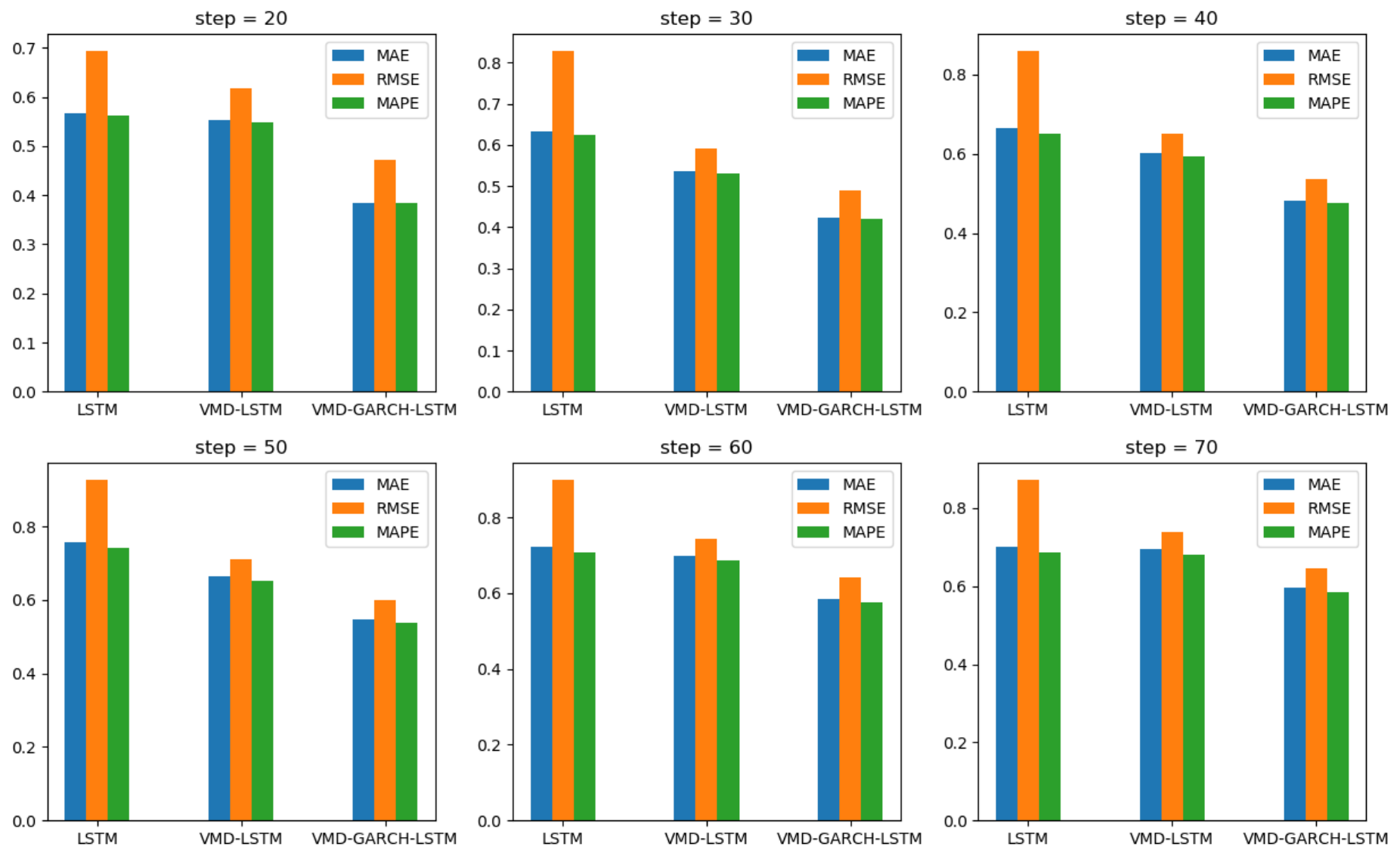}
    \caption{Model Evaluation Metrics under Different Number of Prediction Steps}
    \end{figure}

\begin{table}[!h]
    \caption{Performance of different methods under RMSE MAE and MAPE}
    \centering
    \small
    \begin{tabular}{c c c c c c c c}
        \hline
        Model & Criteria/step & 20 & 30  & 40 & 50 & 60 & 70\\
        \hline
        
        \vspace{1em}
        LSTM  & MAE & 0.567 & 0.633 &  0.666 & 0.758 & 0.722 & 0.701 \\
        \vspace{1em}
        
        & RMSE & 0.693 & 0.828 &  0.859 & 0.926 & 0.898 & 0.871  \\
        \vspace{1em}
        
         & MAPE & 0.562\% & 0.624\% &  0.652\% & 0.742\% & 0.707\% & 0.685\%  \\
        \vspace{1em}
        
        VMD-LSTM  & MAE & 0.553 & 0.537 &  0.603 & 0.663 & 0.697 & 0.695 \\
        \vspace{1em}
        
        & RMSE & 0.617 & 0.591 &  0.650 & 0.710 & 0.744 & 0.737  \\
        \vspace{1em}
        
         & MAPE & 0.549\% & 0.532\% &  0.593\% & 0.651\% & 0.686\% & 0.681\%  \\
        \vspace{1em}

        VMD-GARCH-LSTM  & MAE & 0.385 & 0.424 &  0.483 & 0.547 & 0.586 & 0.597 \\
        \vspace{1em}
        
        & RMSE & 0.472 & 0.488 &  0.537 & 0.601 & 0.641 & 0.645  \\
        \vspace{1em}
        
         & MAPE & 0.383\% & 0.420\% &  0.475\% & 0.537\% & 0.576\% & 0.585\%  \\
        \hline
    \end{tabular}
\end{table}

Figure 10 illustrates the performance of different models under a 70-step prediction scenario. As evidenced, with an increased number of prediction steps, a slight degradation in prediction performance is observed across all models. The performance of LSTM is consistently suboptimal, which highlights the limitations of individual predictive models relative to decomposed-ensemble models. Notably, LSTM appears to be highly sensitive to changes in the number of prediction steps. As shown in Table 3 and Figure 11, a comparison of evaluation indicators reveals that the VMD-LSTM-GARCH model surpasses the other two models in terms of predictive prowess. By leveraging decomposition ensemble techniques, econometric models and neural network methods to optimize forecast performance, our time series forecasting framework produces superior forecast accuracy at various forecast steps. This emphasizes its versatility and adaptability to different prediction models and affirms its robustness.

\section{Conclusion}

In time series forecasting, the predictive errors associated with individual models typically exceed those of comprehensive decomposition-integration models. Decomposition-integration methods prove to be reliable in time series forecasting, while nonlinear ensemble algorithms further improve the forecasting performance of multiscale models. Notably, the integration of GARCH model and neural networks has been identified as yielding substantial improvements. In this study, we propose a novel hybrid time series forecasting framework based on the decomposition-ensemble strategy \citep{yu2015decomposition}. Our framework employs econometric or neural network forecasting methods based on mode-pattern characteristics. Considering the obvious upward trend and sharp fluctuations of most current time series data, our framework can better adapt to the current data characteristics and improve forecasting accuracy. Despite these advancements, opportunities for further exploration remain. In our methodology, the aggregation of predictive outputs from each sub-mode is executed through a rudimentary summation process. It should be noted that employing an independently trained neural network for this specific step of amalgamating results, could potentially lead to divergent outcomes. The performance of the model is also susceptible to the type of decomposition method applied to the time series. These nuances present avenues for future research to investigate and optimize the forecasting process.

\section{Acknowledgement}

The work is supported by the National Natural Science Foundation of China (12371279).

\bibliographystyle{elsarticle-harv}
 
\bibliography{mybib}

\end{document}